# A Large Language Model for Chemistry and Retrosynthesis Predictions

Yueqing Zhang[1#], Wentao Liu[2,3#], Yan Zhang[2,3#], Danyang Xiong[1], Jihang Zhai[1], Hao Hao[2,3], Yu-Cheng Gu[4], Hai-Bo Yang[1], Shuanhu Gao[1], Lianrui Hu[1*], Aimin Zhou[2,3*], Xiao He[1,5,6*]

[1]Shanghai Engineering Research Center of Molecular Therapeutics and New Drug Development, Shanghai Frontiers Science Center of Molecule Intelligent Syntheses, School of Chemistry and Molecular Engineering, East China Normal University, Shanghai, 200062, China

[2]School of Computer Science and Technology, East China Normal University, Shanghai, 200062, China

[3]Shanghai Institute of Artificial Intelligence for Education, East China Normal University, Shanghai, 200062, China

[4]Syngenta Jealott's Hill International Research Centre Bracknell, Berkshire, RG42 6EY, UK

[5]New York University–East China Normal University Center for Computational Chemistry, New York University Shanghai, Shanghai, 200062, China

[6]Chongqing Key Laboratory of Precision Optics, Chongqing Institute of East China Normal University, Chongqing, 401120, China

[*]E-mails: lrhu@chem.ecnu.edu.cn, amzhou@cs.ecnu.edu.cn, xiaohe@phy.ecnu.edu.cn

## Abstract

Large language models (LLMs) have achieved impressive progress across a broad range of general-purpose tasks, but their effectiveness in chemistry remains limited due to scarce domain-specific datasets, and the demand for precise symbolic and structural reasoning. Here we introduce ECNU-ChemGPT(name after East China Normal University), a chemistry-specialized LLMs engineered for deep chemical knowledge understanding and accurate retrosynthetic route planning. Our approach is distinguished by four key strategies: structured prompt-based knowledge distillation from authoritative chemistry textbooks to construct a high-quality question-answering dataset; domain-specific prompt engineering using curated chemical keywords, combined with LLMs APIs for data derivation and knowledge distillation; large-scale fine-tuning on a meticulously cleaned and enriched Pistachio reaction dataset to enhance retrosynthesis prediction accuracy; and integration of BrainGPT, a dynamic multi-model scheduling framework that enables task-specific invocation of multiple specialized models trained for diverse chemistry-related tasks. ECNU-ChemGPT exhibits superior performance on chemistry question-answering and retrosynthetic planning benchmarks, outperforming leading general-purpose models—including Deepseek-R1, Qwen-2.5, and GPT-4o. In retrosynthesis, it achieves a Top-1 accuracy of 68.3% on the USPTO_50K dataset and successfully reconstructed 13 complete experimental pathways for real-world drug molecules from medicinal chemistry journals. These results underscore the effectiveness of domain-adapted fine-tuning combined with dynamic multi-model task scheduling, providing a scalable and robust solution for chemical knowledge question answering, and retrosynthetic planning.

## Introduction

Large language models (LLMs) have revolutionized artificial intelligence (AI) by demonstrating remarkable proficiency in a wide array of fields, particularly excelling in natural language processing (NLP), text generation, and complex reasoning[1-6]. LLMs are typically based on the Transformer architecture, and can process and generate natural language data through unsupervised or self-supervised learning on large-scale datasets. Prominent models in use today include GPT[7], BERT[8], T5[9], and LLaMA[10] etc. After pre-training on vast amounts of text, these models can achieve exceptional performance in a wide range of NLP tasks, such as information extraction, question answering, text generation, and machine translation[11-13].

Despite these impressive achievements in general AI tasks, their efficacy in chemistry—a discipline demanding precise structural interpretation and multi-step logical reasoning—remains critically underexplored[14-17]. While general-purpose LLMs like GPT-4 excel in broad text-generation tasks, they consistently underperform in chemistry-specific applications such as de novo retrosynthetic planning[18,19]. This performance gap stems from four fundamental challenges: (1) Chemistry knowledge representation: chemical information encompasses multiple modalities, including textual descriptions, structural formulas, and specialized notations like Simplified Molecular Input Line Entry System (SMILES)[20], etc., which makes processing multiple formats of data difficult for LLMs[21-23]. (2) Reaction data limitations: traditional retrosynthetic approaches depend on expert-curated rules or limited template libraries, while data-driven, LLM-based methods face challenges due to insufficient reaction diversity and molecule representation error[24]. (3) Hallucinations and factual errors in chemical knowledge: although LLMs have demonstrated the ability to outperform many chemistry experts in overall evaluations, they still exhibit notable shortcomings on fundamental questions and frequently generate overly confident yet incorrect responses[25-27]. This suggests that while current LLMs possess strong capabilities in memorizing and reasoning over chemical information—allowing them to address a broad range of common and integrative problems—they lack robust verification mechanisms, resulting in hallucinations. (4) Single-model limitations in multi-task applications: optimizing one model to perform well across diverse tasks, such as retrosynthetic planning, knowledge-based question answering, pose a significant challenge[28-31].

The practical implications of these challenges are starkly evident in various benchmark evaluations. Herein, we introduce ECNU-ChemGPT, an LLMs specifically tailored for chemical applications (as shown in Fig 1). It begins with the construction of diverse training datasets, including a knowledge-based question-answer pairs derived from textbooks and keyword-guided expansions, a manually labeled multi-model question answering dataset, and a cleaned reaction dataset extracted from the Pistachio database[32]. These datasets were used to fine-tune task-specific base models, with different models trained for different tasks. To facilitate the use of multiple models, BrainGPT is employed to schedule and manage them. The BrainGPT model serves as a central scheduler, intelligently selecting and coordinating the appropriate sub-models based on the task type. This enables ECNU-ChemGPT to support a wide range of chemistry-related applications—such as question answering, reaction prediction, and molecular property analysis—through a unified and extensible framework. ECNU-ChemGPT addresses the challenges above through three strategies: 1) Knowledge distillation through structured prompts: chemical concepts extracted from textbooks are systematically converted into a structured question-answer format using prompt engineering. This approach, combined with knowledge distillation techniques, significantly enhances the accuracy and depth of domain-specific question answering. 2) Keyword-guided data expansion and derivation: Keyword-guided data expansion and derivation: to reduce hallucinations of large language models in the chemistry domain and enhance data diversity, ECNU-ChemGPT adopts a keyword-based derivation and expansion strategy, leveraging carefully selected chemical keywords and LLM-driven generation to construct a comprehensive, high-quality dataset. 3) Enhanced retrosynthetic route planning: by fine-tuning the LLMs using larger (about 3.7 million chemical reactions), high-quality chemical reaction datasets, ECNU-ChemGPT substantially improves the accuracy and reliability of the prediction of retrosynthetic routes. 4) Dynamic multi-model task scheduling: ECNU-ChemGPT integrates the BrainGPT module, which dynamically invokes specialized sub-models (e.g., question-answering, retrosynthetic planning, etc.) based on the input query, thereby facilitating seamless multifunctional integration.

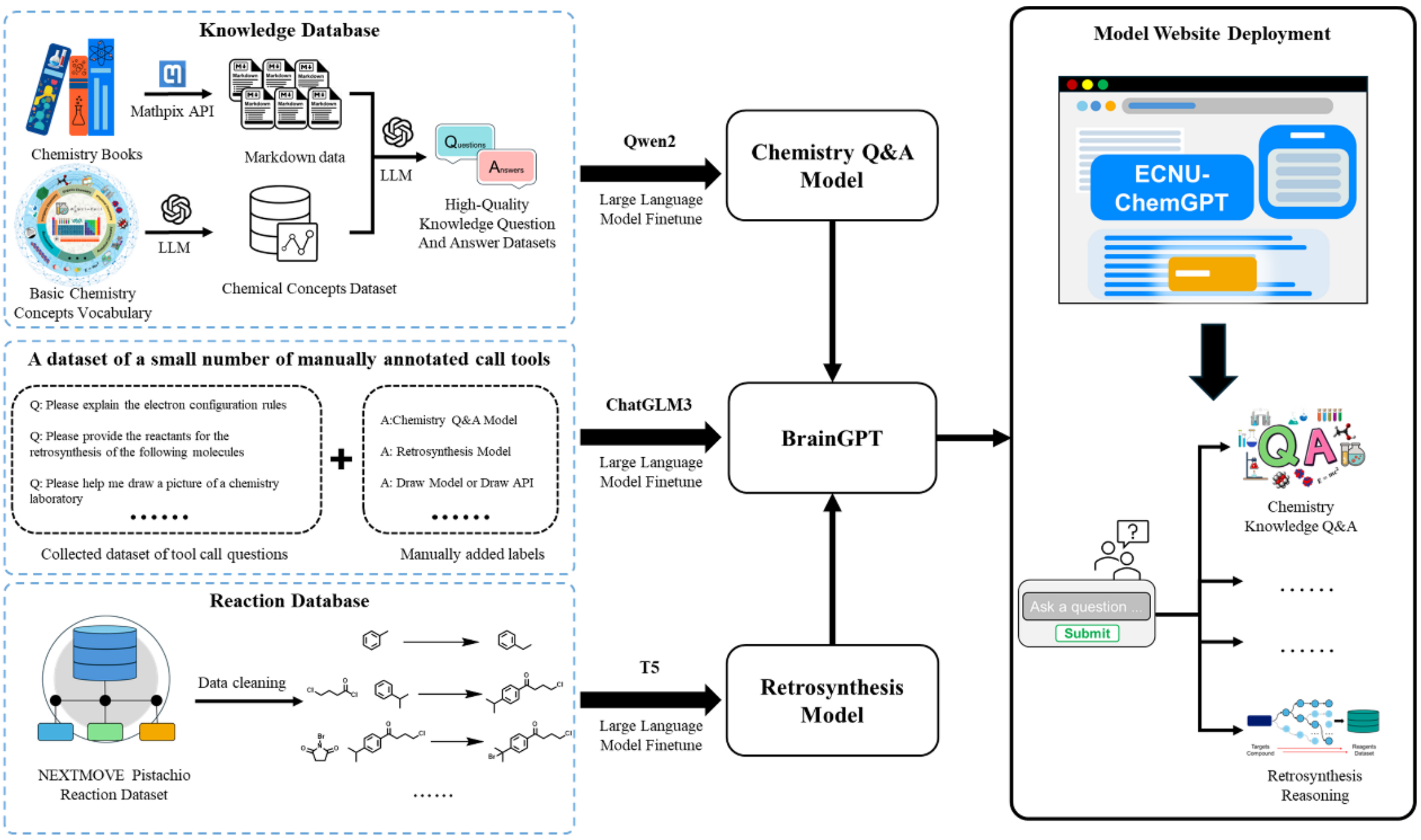

**Figure 1. | Architecture and workflow of the ECNU-ChemGPT model.** The left side of the diagram illustrates the training datasets, including a knowledge-based question-answering dataset constructed from textbooks content and keyword-guided expansion, a manually labeled multi-model question answering dataset and a cleaned reaction dataset derived from the Pistachio database. These datasets are used to fine-tune corresponding base models, which are subsequently integrated into the final ECNU-ChemGPT framework, accessible via a unified web interface (https://chemgpt.lrcwtech.com).

## Results

### Chemistry question-answer benchmark testing

Although many LLMs in chemistry have been developed and deployed, a significant challenge persists: these models often struggle to deliver accurate answers to chemistry-related queries and are prone to common-sense errors[33,34]. To evaluate the effectiveness of the model fine-tuning with supplementary chemical data, benchmark tests were conducted and compared with the base model. The results are summarized in Figure 2a.

The C-Eval dataset[35] is a widely recognized benchmark in Chinese natural language processing (NLP), providing a standardized platform for evaluating model performance across multiple tasks, including text generation, classification, sentiment analysis, reading comprehension, and reasoning. In this study, the evaluation focused on the chemistry sections of C-Eval, specifically the junior high school, high school, and university-level chemistry questions. These sections contain numerous LaTeX formulas and require advanced reasoning skills, making them particularly valuable for assessing a model's ability to handle complex chemical problems. To effects of fine-tuning, performance was compared against the baseline models, such as Qwen2-72B-instruct[36], GPT-4o[7], Deepseek-R1[37]. Remarkably, the fine-tuned model demonstrated better performance across all chemistry levels and achieved a score of 83.33 in university-level chemistry， which represented a substantial improvement over the original Qwen2 model, which scored 79.17. Furthermore, evaluation was conducted using the C-MHChem-Benchmark-Chinese-Middle-High-School-Chemistry-Test[38], a high-quality, single-choice benchmark comprising 600 questions meticulously curated from official Chinese middle and high schools chemistry exams over the past 25 years. Designed to assess a model's ability to understand and answer chemistry-related queries at different educational levels, this benchmark provided a rigorous assessment of domain-specific generalization. The fine-tuned model ranked second only to Deepseek-R1[37] in accuracy and reasoning performance, and outperformed all other models with similar sizes of parameters. These results underscore the effectiveness of integrating domain-specific knowledge with guided data expansion through a systematic strategy—extracting authoritative textbook content, generating instruction-based prompts, and scaling them into a high-quality dialogue dataset—to enhance model capabilities in specialized scientific domains.

We further assess the model's general reasoning ability using the GSM8K dataset[39], which was developed by OpenAI and contains 8.5K high-quality, linguistically diverse elementary school math word problems. Given that the fine-tuning process focused exclusively on chemistry, performance gains on GSM8K were expected to be minimal, and the results confirmed this, showing a marginal improvement over the base models. This suggests that while fine-tuning with domain-specific data enhances task-specific performance, it does not necessarily generalize to broader reasoning capabilities. Nevertheless, the model retained its original reasoning ability on GSM8K, indicating that fine-tuning chemistry data did not compromise its general problem-solving capacity. These findings align with previous research suggesting that targeted fine-tuning primarily improves domain-specific performance without broadening general reasoning ability[40-42].

To evaluate the model's ability on open-ended questions, especially those involving chemical formula derivation, we employed a college entrance examination dataset in the OpenCompass[43]. As these questions involve subjective content, two relatively authoritative LLMs were employed for evaluation. The generated responses were assessed according to four criteria: chemical foundations reliability (CFR), chemical knowledge accuracy (CKA), answer accuracy (AA), and language and formula expression accuracy (LFEA). Each criterion had a maximum score of 25 points, for a total of 100. Figure 2b presents the scores of ChatGPT-4o and Deepseek-R1 across each criterion, along with the total scores. The fine-tuned ECNU-ChemGPT outperformed its pre-fine-tuning version in most categories, indicating that the addition of the two datasets improved chemical formula derivation, logical reasoning, and the level of chemical knowledge. As shown in Fig 2b, the average scores across all abilities are higher than those of the base model, and the overall score is also improved. Figure 2c and 2d demonstrate that ECNU-ChemGPT outperforms the base model on two different datasets across the score distribution ranges of both scoring models. In contrast, the base model exhibits a few outliers, suggesting insufficient accuracy in question answering. An example of a question from the Opencompass dataset, along with the corresponding model-generated answer, is shown in Supplementary Fig S1 and S2. By comparing the answers of ECNU-ChemGPT and Qwen2-72B-instruct, the accuracy of chemical knowledge, understanding of chemical principles, and logical ability has all improved. Overall, these results demonstrate that incorporating textbook-derived data and keyword-guided data expansion significantly enhances the model's reasoning capabilities in complex, domain-specific chemistry tasks.

**a**

| models | C-Eval College chemistry | C-Eval High school chemistry | C-Eval Middle school chemistry | MMLU College chemistry | MMLU High school chemistry | C_MHChem | GSM8K | Average |
|---|---|---|---|---|---|---|---|---|
| Llama3_1-8b-instruct | 41.67 | 33.33 | 83.33 | 51.67 | 71.59 | 52.32 | 83.35 | 59.61 |
| Internlm2_5-7b-chat | 55.56 | 75.44 | 95.00 | 50.33 | 62.89 | 47.34 | 82.97 | 67.07 |
| GPT3.5 | 33.33 | 49.12 | 81.67 | 51.33 | 53.20 | 50.82 | 75.29 | 56.39 |
| GPT-4o-2024-05-13 | 66.67 | 52.63 | 95.00 | 70.67 | 89.49 | 62.98 | 89.56 | 75.29 |
| GLM-4-9b-chat | 38.89 | 61.40 | 98.33 | 38.67 | 52.55 | 67.97 | 59.62 | 59.63 |
| Mistral-7b-instruct-v0.3 | 29.17 | 36.84 | 56.67 | 33.00 | 33.99 | 32.86 | 53.80 | 39.48 |
| ChemLLM | 27.78 | 45.61 | 85.00 | 36.67 | 38.92 | 47.95 | 41.72 | 46.24 |
| Deepseek-V3 | 77.78 | 77.19 | 95.00 | 66.33 | 85.22 | 74.52 | 90.88 | 80.99 |
| Deepseek-R1 | 47.22 | 63.16 | 96.67 | **71.33** | **91.79** | **83.81** | 92.65 | 78.09 |
| Qwen2.5-7b-instruct | 66.67 | 73.68 | 95.00 | 54.33 | 75.37 | 63.39 | 80.44 | 72.70 |
| Qwen2-72b-instruct | 79.17 | **89.47** | **100.00** | 60.33 | 76.85 | 75.34 | 90.12 | 81.61 |
| ECNU-ChemGPT | **83.33** | **89.47** | **100.00** | 61.00 | 77.34 | 82.45 | **92.93** | **83.79** |

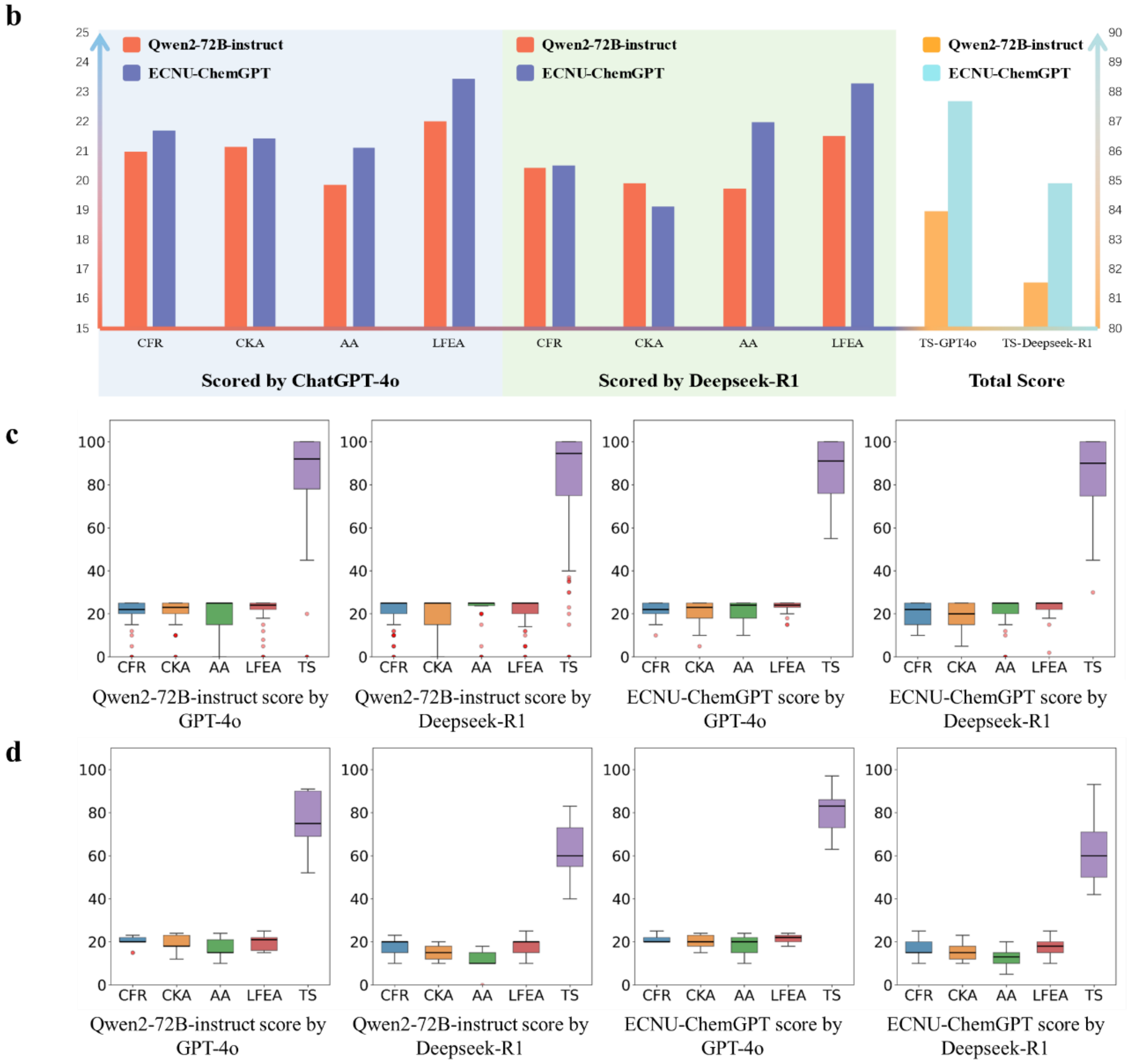

**Figure 2. | Benchmark evaluation of ECNU-ChemGPT and baseline models. a,** Benchmark scores of various models across different datasets. **b,c,d** Performance comparison between Qwen2-72B-instruct and ECNU-ChemGPT on multiple chemistry-related tasks. Evaluation metrics include CFR (chemical formula recognition), CKA (chemical knowledge assessment), AA (answer accuracy), LFEA (long-form answer assessment) and TS (task score), with scores provided by two

independent LLMs evaluators: ChatGPT-4o and Deepseek-R1. **b.** Comparison of average ability scores between ECNU-ChemGPT and Qwen2-72B-instruct. ECNU-ChemGPT achieves consistently higher scores across all evaluated abilities, along with an improved overall score. **c,d**. Performance comparison between ECNU-ChemGPT and Qwen2-72B-instruct on two different datasets. ECNU-ChemGPT demonstrates superior score distributions in both scoring models, while the base model exhibits a few of outliers, indicating lower accuracy and consistency in question answering.

**Retrosynthetic planning**

The model's performance on retrosynthesis task was first assessed using two benchmark datasets. Since the USPTO_50K dataset[44] was not include in the training process and the data in it was not in the training set, the entire dataset was employed as the test set. The evaluation metric was Top-1 accuracy, defined as the proportion of cases in which the model's top-ranked prediction exactly matched the ground-truth reaction. ECNU-ChemGPT was compared with several general-purpose models, including ChatGPT-4o-mini, ChatGPT-o3-mini, Deepseek-R1, and Deepseek-V3, as well as the chemistry-specialized LLMs ChemDFM[45]. On the USPTO_50K dataset, ECNU-ChemGPT achieved excellent performance in both settings, with or without the reaction type provided, reaching a Top-1 accuracy of 68.3% when the reaction type was known. As shown in Figure 3a, ECNU-ChemGPT consistently outperformed all baseline models across evaluation conditions.

To further evaluate retrosynthetic reasoning capabilities, evaluations were extended to real-world drug synthesis routes developed over the past decade. A total of 262 complete synthesis routes, comprising 1,448 individual reactions, were collected from the *Journal of Medicinal Chemistry*. These extracted reaction data are now publicly available on GitHub. On this literature-derived reaction dataset, ECNU-ChemGPT continued to outperform all baseline models. The distribution of reaction types in both the training and the collected evaluation datasets is shown in Figure 3b, and the prediction results of ECNU-ChemGPT are summarized in Figure 3c. Among the 1,448 single-step reactions, 1,424 yielded valid predictions, with 884 accurately matching the experimental outcomes. For the 262 full retrosynthetic routes, 242 valid predictions were obtained, 13 of which were exactly consistent with the literature-reported synthetic routes. When exact matches were required, the Top-1, Top-3, and Top-10 accuracies reached 19.6%, 45.7%, and 64.6%, respectively (Figure 3c). Considering the inherent non-uniqueness of synthetic routes and the frequent presence of substituent variations (e.g., chlorobenzene vs. bromobenzene) that do not affect the overall outcome, additional evaluation was conducted based on molecular similarity. Under this relaxed criterion, the Top-1, Top-3, Top-10, and Top-15 accuracies improved to 20.9%, 50.8%, 85.1%, and 98.2%, respectively, as shown in Supplementary Fig S3. When comparing the model's performance on the USPTO-50K dataset and the collected reaction dataset from the medicinal chemistry journal, a significantly lower TOP-1 accuracy is observed on the latter. We speculate that this discrepancy arises from the fact that reactions in USPTO-50K are general organic synthesis reactions, whereas drug synthesis requires more stringent considerations, including safety, stability, and cost-effectiveness. Moreover, the reactions selected for actual drug synthesis are not always the optimal routes suggested by retrosynthetic analysis, which further contributes to the variation in TOP-1 accuracy between the two datasets. These results demonstrate the strong generalization ability of ECNU-ChemGPT in retrosynthetic planning. Manual inspection further confirmed that many predicted routes, though different from the experimentally reported ones, were chemically reasonable, highlighting the model's ability to propose alternative yet viable synthetic pathways.

An automated interface for multi-step retrosynthesis has been developed and is publicly available at https://multiasr.lrcwtech.com/, enabling route generation for diverse target compounds. Figure 3d presents six representative examples from the set of 13 previous identified predicted synthesis pathways that exactly matched those reported in the *Journal of Medicinal Chemistry*, with additional results shown in Supplementary Fig S4. In the text, routes 1 to 6 are the synthetic routes of Tolvaptan[46], Remdesivir[47], Daridorexant[48], Abemaciclib[49], Baricitinib[50], and Betrixaban[51]. Here, a brief analysis is given for the first synthetic route, the synthesis of Tolvaptan. First, the benzylic hydroxyl group of tolvaptan can be introduced by reducing the corresponding carbonyl compound, which in turn can be constructed via an amidation reaction between aniline and o-formyl chloride, with the aniline precursor obtained through the reduction of a nitro compound. Further retrosynthetic analysis revealed that the nitrogen-containing benzazepine ring structure can be cleaved via amide bond scission into p-nitrobenzoyl chloride and a secondary amine compound. The latter can be obtained through Ts-deprotection and $\alpha$-decarboxylation, while the key seven-membered ring framework can be efficiently constructed via Dieckmann condensation.

a

| Model | USPTO-50K | | Collected reaction dataset |
|---|---|---|---|
| | Reaction class known(TOP-1 Accuracy) | Reaction class unknown(TOP-1 Accuracy) | Reaction class unknown(TOP-1 Accuracy) |
| ChatGPT-4omini | 0.008 | 0.002 | 0.069 |
| ChatGPT-o3mini | 0.446 | 1.737 | 1.174 |
| Deepseek-V3 | 0.384 | 0.326 | 2.141 |
| Deepseek-R1 | 1.280 | 1.332 | 2.072 |
| ChemDFM-8B | 4.380 | 1.560 | 0.620 |
| ChemDFM-13B | 5.290 | 2.320 | 0.550 |
| ECNU-ChemGPT | **68.30** | **50.10** | **19.61** |

b

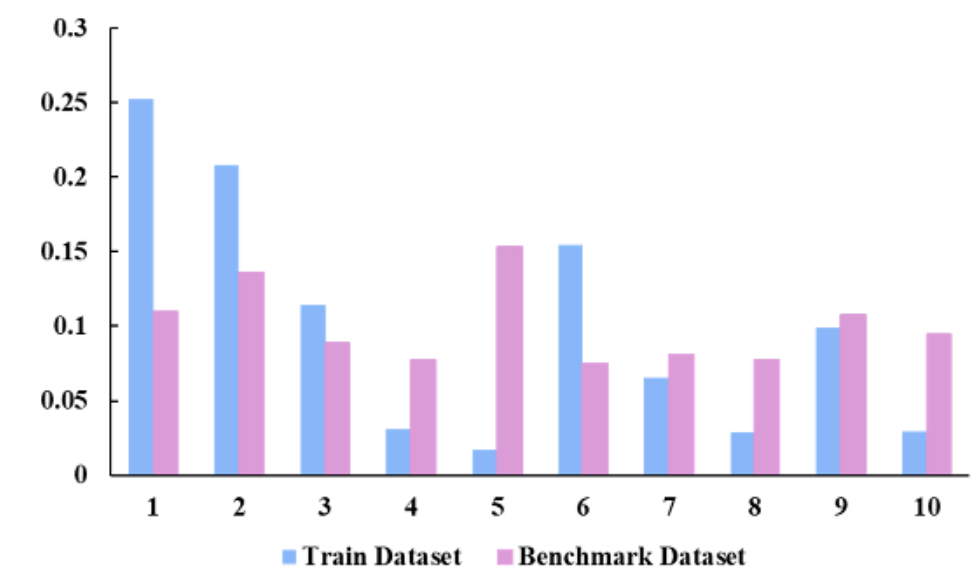

c

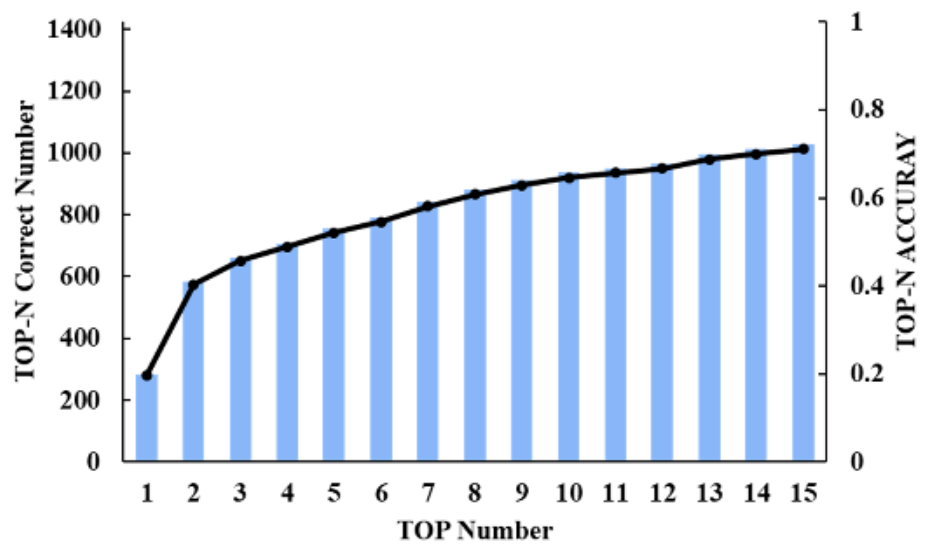

d

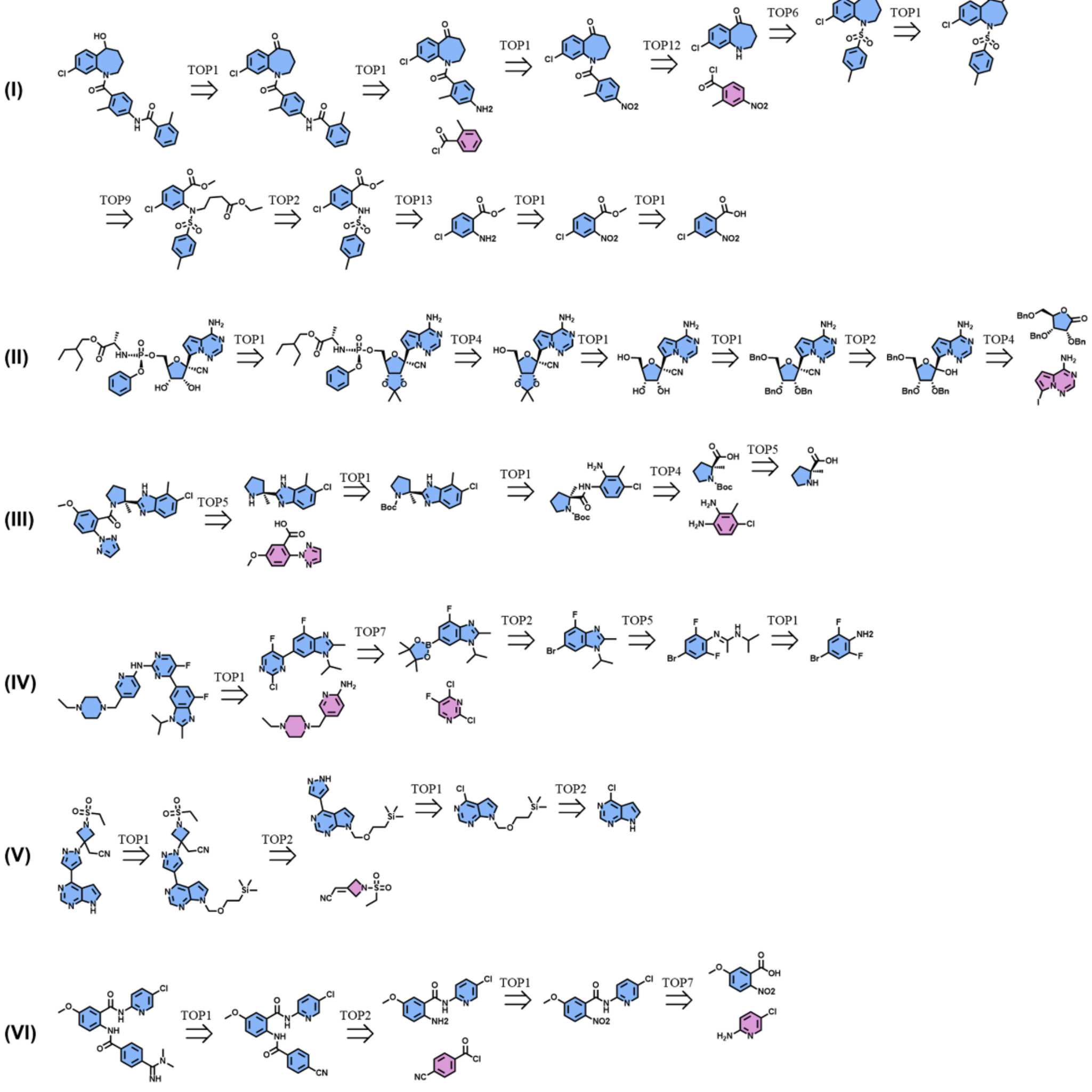

**Figure 3. | Performance of ECNU-ChemGPT and baseline models on retrosynthetic planning benchmarks and representative prediction cases. a**, TOP-1 accuracy comparison of ChatGPT-4omini, Deepseek series, ChemDFM series and ECNU-ChemGPT on the USPTO_50K dataset and collected reaction datasets,

summarized in tabular form. **b**, Distribution of reaction types in the training set and the collected real-world reaction dataset. **c**, Prediction performance of ECNU-ChemGPT across different TOP-N thresholds. As N increases, both the number of correct predictions and overall accuracy improve, reaching 85.1% at TOP-10 without reaction type. **d**, Retrosynthetic pathways predicted for six real-world drug molecules. Blue highlights indicate the model-predicted product at each step, while purple denotes the additional reactants required in the corresponding reactions.

## Discussion

In this work, we developed ECNU-ChemGPT, a large-scale language model tailored specifically for the chemistry domain, designed to address critical challenges in chemical knowledge question answering and retrosynthetic planning. By leveraging textbook-derived knowledge distillation, structured prompt engineering, and instruction fine-tuning, the model demonstrates enhanced capability to comprehend and generate accurate responses to complex chemical queries. The retrosynthetic route planning framework further demonstrated the model's performance through the validation of drug synthesis route design, achieving improved reliability in generating synthetic routes. To enhance the adaptability and integration of multiple tasks, we introduced BrainGPT, a multi-model scheduling model that dynamically calls specialized models for reasoning. This architecture provides a scalable and customizable AI-driven solution for diverse chemical research scenarios. Benchmark evaluations confirmed that ECNU-ChemGPT surpassed general LLMs, such as GPT-4o, Deepseek-V3, in domain-specific large language model for chemistry, especially in retrosynthetic reasoning and chemistry question answering. It outperformed other general models on multiple benchmark datasets. Compared with the ChatGPT-4o model, the average score of each evaluation increased by 8.5%, compared with Deepseek-R1, the average score increased by 5.7 %, and compared with the base model, the average score increased by 2.18 %. Although the improvement in knowledge question answering is only moderate compared to the basic model, its advantage in retrosynthesis planning is still relatively large compared to other models. These advances underscore the importance of domain-specific fine-tuning in advancing LLMs for scientific applications and provide new strategies for retrosynthetic route planning. Our work highlights the potential of fine-tuning domain-specific large language models for chemistry. Future efforts will focus on expanding high-quality chemical datasets, improving retrosynthetic accuracy, and enhancing model generalizability across a wide range of chemical applications. With continued development, ECNU-ChemGPT is expected to become a powerful tool in various areas of chemistry, including drug discovery, retrosynthetic route planning, and beyond.

## Methods

To overcome the challenges encountered by LLMs in the field of chemistry, an integrated model has been established, incorporating four core strategies: knowledge distillation through structured prompts, keyword-guided data expansion and derivation, enhanced retrosynthetic route planning using a large dataset, and dynamic multi-model task scheduling using BrainGPT.

### Base model

To enhance the model's reasoning capabilities and its ability to extract and process high-dimensional features from raw data, a performance comparison was conducted across various general-domain LLMs. On benchmarks such as MMLU[52], GSM8K[39], and C-Eval[35], Qwen2 exhibited improved performance compared to Qwen1.5, with higher scores across evaluations on some datasets( see Supplementary Table S1). Compared to LLaMA, Qwen2 achieved better performance in both Chinese and English question answering, although a slightly lower score was observed in coding ability. Given that the model is intended for chemical knowledge question answering, Qwen2 was selected as the base model for chemistry-specific LLM-based reasoning. The Qwen2-72B-Instruct model was fine-tuned using the parameter-efficient LoRA (Low-Rank Adaptation)[53] technique. For fine-tuning parameter settings, see Supplementary Table S2 and Table S3. Under this configuration, a total of 862,650,368 trainable parameters were introduced by LoRA, accounting for approximately 1.17% of the full model's 73,568,854,016 parameters. This setup was adopted to significantly reduce the computational cost of fine-tuning while preserving the original model's generation capability and knowledge representation performance.

In the field of retrosynthetic planning, existing models with high reported predictive accuracy were surveyed, revealing three main categories of template-free approaches: fingerprint-based[54], sequence-based[55], and graph-based[56,57]. Despite achieving high accuracy on benchmark datasets such as USPTO_50K, some routes inferred by the model sometimes fail to synthesize the current molecules in experiments. In contrast, LLMs provide a promising template-free alternative by offering strong generalization capabilities, multi-step logical reasoning, high interactivity, and scalability. These features make them well-suited for complex retrosynthetic route prediction. Consequently, leveraging LLMs for retrosynthetic planning inference tasks was established as one of the primary goals for the development of this model. Considering the trade-off between model size, computational cost, and predictive performance, we opted not to use excessively large models, as prior studies have shown that moderate-sized models can achieve competitive results in chemical reaction prediction tasks with significantly lower resource requirements. T5-large was selected as the base model for retrosynthetic route prediction due to its favorable balance between efficiency and accuracy[58,59].

## Datasets

### Chemistry knowledge question and answer datasets from books

As a fundamental natural science, chemistry explores the composition, structure, properties, and transformations of matter, spanning scales from atomic and molecular reactions to macroscopic phenomena. This intrinsic complexity presents significant challenges for LLMs, particularly in accurately handling chemical formulas, specialized terminology, and symbolic representations. Such issues can severely limit the effectiveness of LLMs in chemistry-related tasks, where precision is critical. To address these challenges and construct a high-quality chemical dataset, a systematic strategy was adopted (Figure 4a). First, chemical knowledge was extracted from over 400 authoritative Chinese and English chemistry-related textbooks, covering disciplines including organic, inorganic, molecular, physical, and biochemistry, etc.. This foundational corpus provided detailed knowledge of chemical reactions, molecular structures, and fundamental chemical principles. Based on the extracted content, instruction-based prompts were generated to guide model responses appropriately to chemistry-related queries, thereby improving the model's ability to process complex concepts with greater accuracy. Finally, the instruction fine-tuning data were leveraged to build a high-quality dialogue dataset containing 848,176 question–answer pairs, as shown in Figure 4b, designed to equip the model with both comprehensive chemical understanding and the capacity for precise, contextually appropriate dialogue across a wide range of chemical topics. Through this process, a robust chemistry-specific dataset was established, enabling fine-tuning of the model to enhance its accuracy and effectiveness in chemical knowledge representation and reasoning. Supplementary Fig S5 shows the process of extracting textbook knowledge, such as narrative descriptions, tabular data, and chemical formulas, and transforming it into training data using GPT-4o.

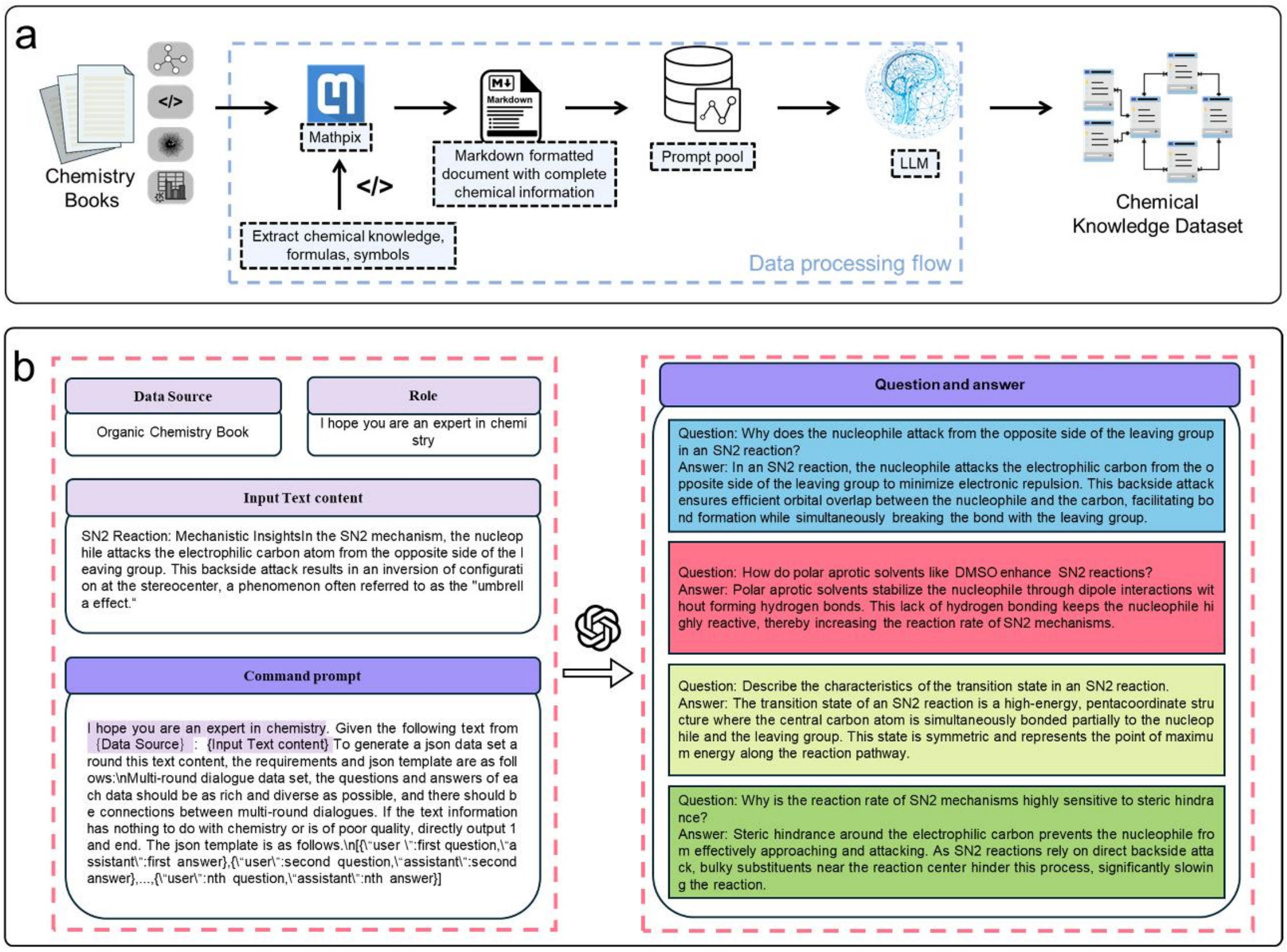

**Figure 4. | Dataset preparation from chemistry books. a**, Workflow for extracting and processing data from chemistry textbooks. **b**, Example of prompt design for generating high-quality dialogue dataset using LLMs.

**Chemistry knowledge question-answer datasets derived via API-based expansion**

Large language models (LLMs) applied to the chemical domain are prone to hallucination, where inaccurate or fabricated content is generated[60,61]. This issue becomes critical in chemistry, a domain that demands high factual precision. Additionally, insufficient data diversity—especially in emerging or interdisciplinary areas—can limit model generalization and reliability. To address these limitations, a keyword-guided expansion framework was developed, integrating keyword-guided data expansion with instructional data derivation. This approach aims to reduce hallucination through grounding in diverse, high-quality domain knowledge and to enhance data diversity by covering a broader spectrum of chemical topics and interdisciplinary contexts. LLMs possess the capacity to absorb and reorganize information from large-scale corpora, including up-to-date literature and domain-specific databases. This property allows for the automatic integration of new knowledge, contributing to sustained model relevance across diverse chemical subfields, such as biochemistry, medicinal chemistry, and environmental chemistry. In contrast, traditional sources, such as textbooks, are often outdated and insufficient in representing newly emerging concepts or cross-disciplinary insights, thus limiting their utility in training robust models.

The data derivation process followed three main steps (Figure 5): (1) Collection of basic chemical instructions. A vocabulary set comprising 120 fundamental chemical terms was assembled, covering core areas including inorganic, organic, physical, analytical, and biochemistry etc. These foundational terms served as seeds for subsequent expansion. (2) Generation of a complex instruction pool based on LLMs. The basic instructions were combined with designed prompts and expanded iteratively through multiple rounds of LLM-driven generation, yielding a broad, hierarchical pool of 406,806 instruction fine-tuning entries. To manage data volume, 10% of the expanded pool was randomly selected for subsequent processing. (3) Generation of a high-quality dialogue dataset. The selected instruction entries were further transformed into a high-quality dialogue dataset containing 1,176,000 question-answer pairs. By enhancing both the depth and breadth of chemical knowledge representation, this methodology enables the fine-tuned LLMs to more accurately understand chemistry knowledge. The resulting dataset enhances both concept depth and domain breadth, enabling the fine-tuned model to better in chemistry knowledge, support interdisciplinary reasoning, and reduce hallucinated outputs through exposure to verified and diverse chemical content.

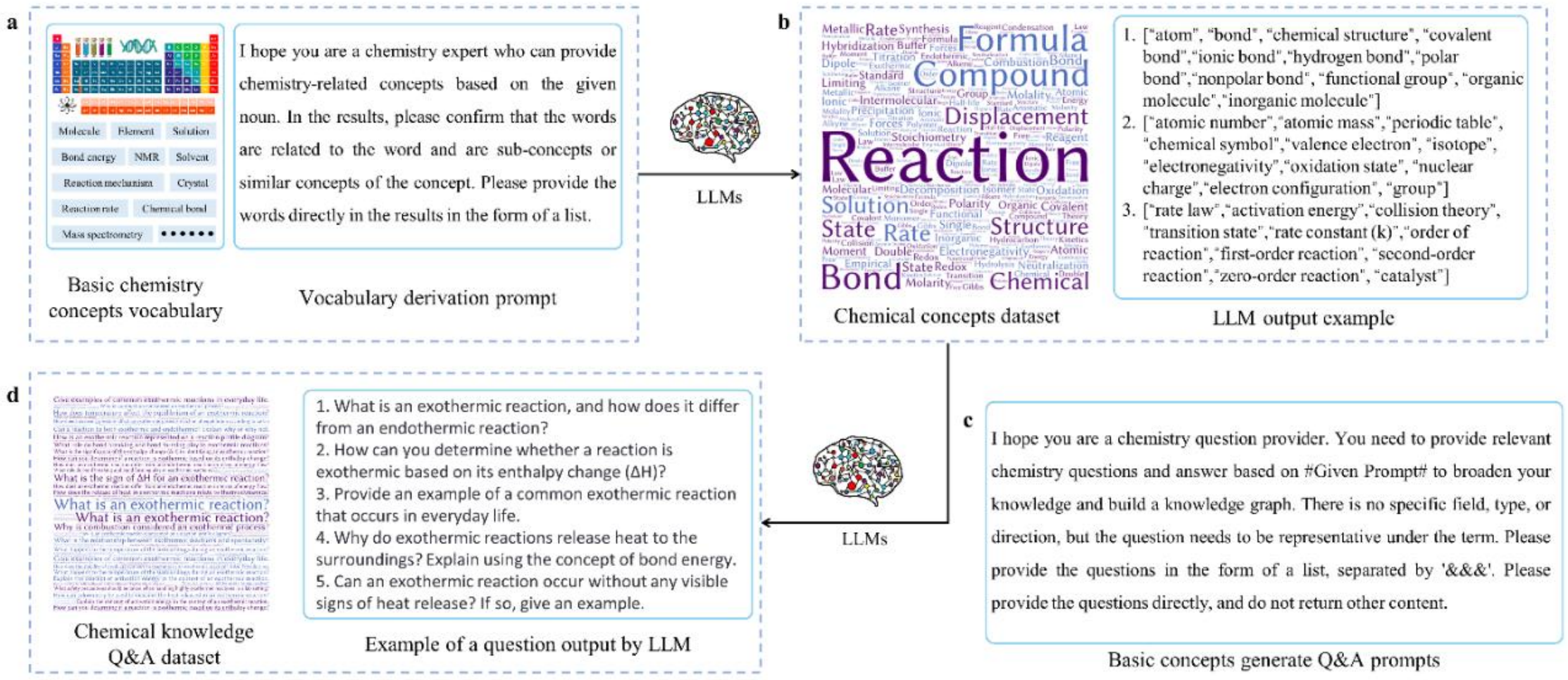

**Figure 5. | Workflow for building a high-quality chemistry question-answer dataset via LLM-driven knowledge expansion.** a. Basic chemistry vocabulary and derivation prompt design: A list of fundamental chemistry terms was curated. A vocabulary derivation prompt is designed to guide the LLMs to generate related sub-concepts associated with each given word. b. Chemical concept expansion and dataset generation: Based on the prompts from (a), LLMs output a rich set of related chemistry terms, forming a chemical concepts dataset. c. Q&A prompt generation based on concepts: Using the expanded concepts from (b), Q&A generation prompts are designed and fed into LLMs. These prompts guide the model to generate chemistry-related questions and answers that deepen the conceptual understanding. d. Construction of a chemistry knowledge Q&A dataset: LLMs generate Q&A pairs based on the prompts in (c).

**Retrosynthesis datasets**

To improve model performance on retrosynthesis planning tasks and broaden its applicability, the USPTO_50K dataset was initially employed. However, limited data volume made it difficult to fine-tune LLMs effectively, resulting in poor learning of molecular representations conveyed by SMILES strings[62]. To address this limitation and improve the performance of the retrosynthetic model, the larger Pistachio dataset (The pistachio database version number is 2021_Q4 and 2024_Q1)[32], comprising 24,550,918 reaction records（about 6 million unique reaction records）, was selected and subsequently cleaned (Figure 6). As shown in Fig. 6a, the dataset includes information such as reaction types, reactant SMILES, product SMILES, catalyst/solvent SMILES, and other relevant fields. A two-step cleaning process was implemented to improve data quality and consistency. First, considering that a single molecule can be represented by multiple valid SMILES strings, RDKit[63] was used to standardize the SMILES representations and remove duplicates (Fig. 6b). Second, reactions involving molecules with a heavy atom count $\leqslant$ 3 or lacking carbon atoms were excluded to eliminate irrelevant or inorganic reactions (Fig. 6c). Finally, reaction records were discarded if the difference in the number of heavy atoms between reactants and products exceeded 50% of the heavy atom count in the reactants. Following these cleaning and deduplication steps, a curated dataset comprising 3,781,569 high-quality reaction records was obtained, providing a robust foundation for fine-tuning the retrosynthesis model.

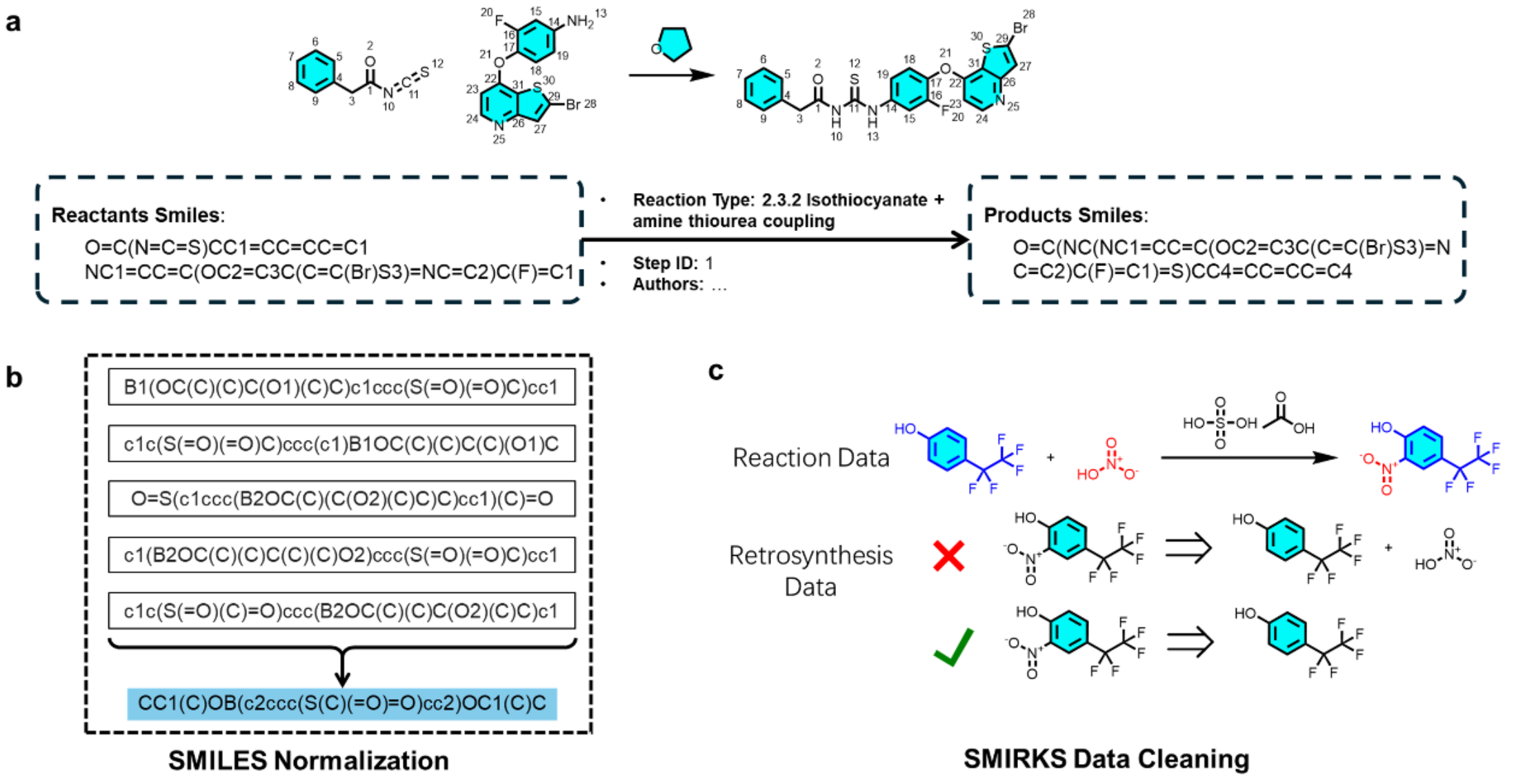

**Figure 6. | Data cleaning and preprocessing of the Pistachio reaction dataset. a**, Example of chemical reaction records extracted from the Pistachio database. **b**, Illustration of SMILES normalization using RDKit to standardize molecular representations. **c**, Example of eliminated reactions involving reagents with fewer than 3 heavy atoms or lacking carbon atoms to remove irrelevant or inorganic reactions during data cleaning.

## Model fine-tuning

After selecting the base models and preparing the datasets, instruction fine-tuned was performed on the Qwen2-72B-Instruct model using the curated chemical knowledge dataset, employing the efficient LoRa parameter fine-tuning method. In parallel, the chemical reaction dataset was used to fine-tune the T5-Large-770M model. In both cases, full LoRA fine-tuning was applied by inserting LoRA layers into the attention and linear layers. The base model architecture and LoRA fine-tuning schematics are shown in Supplementary Fig S6. All model tuning procedures were conducted on 8 NVIDIA A800 80GB GPUs.

## BrainGPT

Currently, a single LLMs often exhibits limited adaptability when addressing multiple tasks[64]. To overcome this, several tool-augmented models have been proposed to integrate multiple tool functionalities. These models primarily rely on static prompts provided at the beginning of each interaction to guide tool usage. However, this strategy often performs poorly for less frequently used tools due to limited training data coverage, and its performance further deteriorates when more than three tools are integrated. To resolve this issue, we developed BrainGPT, a multi-model dialogue model designed for dynamic tool invocation (Figure 7). BrainGPT can effectively solve the problem of multi-model invocation by using manually labelled data and fine-tuning a base model optimized for fluent multi-model dialogue, low deployment overhead, and efficient dynamic tool integration. BrainGPT possesses the following capabilities: (1) General chemistry question answering, through access to external APIs or locally deployed models such as the Qwen-based chemistry question-answer model described in ECNU-ChemGPT. (2) Molecular retrosynthetic reasoning, using retrosynthetic reasoning models, such as the T5 retrosynthetic reasoning model described in ECNU-ChemGPT; (3) Chemical knowledge retrieval, supporting real-time updates from external knowledge databases[65]; (4) General-purpose image generation, based on the stable-diffusion-xl-base-1.0 model[66]. Owing to its flexible architecture, BrainGPT enables seamless tool updating and customized model expansion, providing a scalable solution for complex, multi-task applications in chemistry. If new models or APIs need to be integrated, corresponding invocation Q&A data can be added and used for fine-tuning to enable support for the new tools. Moreover, when an existing model is updated without changes to its functionality, there is no need to retrain—the system can be updated simply by modifying the model path.

BrainGPT is built upon the ChatGLM3-6B[67] architecture and was fine-tuned in two stages to reach its current version. As illustrated in Figure 7, the BrainGPT implementation process consists of two stages. Initially, a small set of tool-calling datasets was manually annotated to train the first version. The questions in this dataset represent commonly used queries for currently supported tools, and the corresponding answers indicate which model should be invoked. In the second stage, common question patterns are collected and analyzed from backend logs. These questions are then manually labeled with the corresponding model tags to construct a new dataset, which is subsequently used for further fine-tuning of the model. Additionally,

BrainGPT adopts the same tool invocation protocol as ChatGLM3, ensuring compatibility and consistency in multi-model integration.

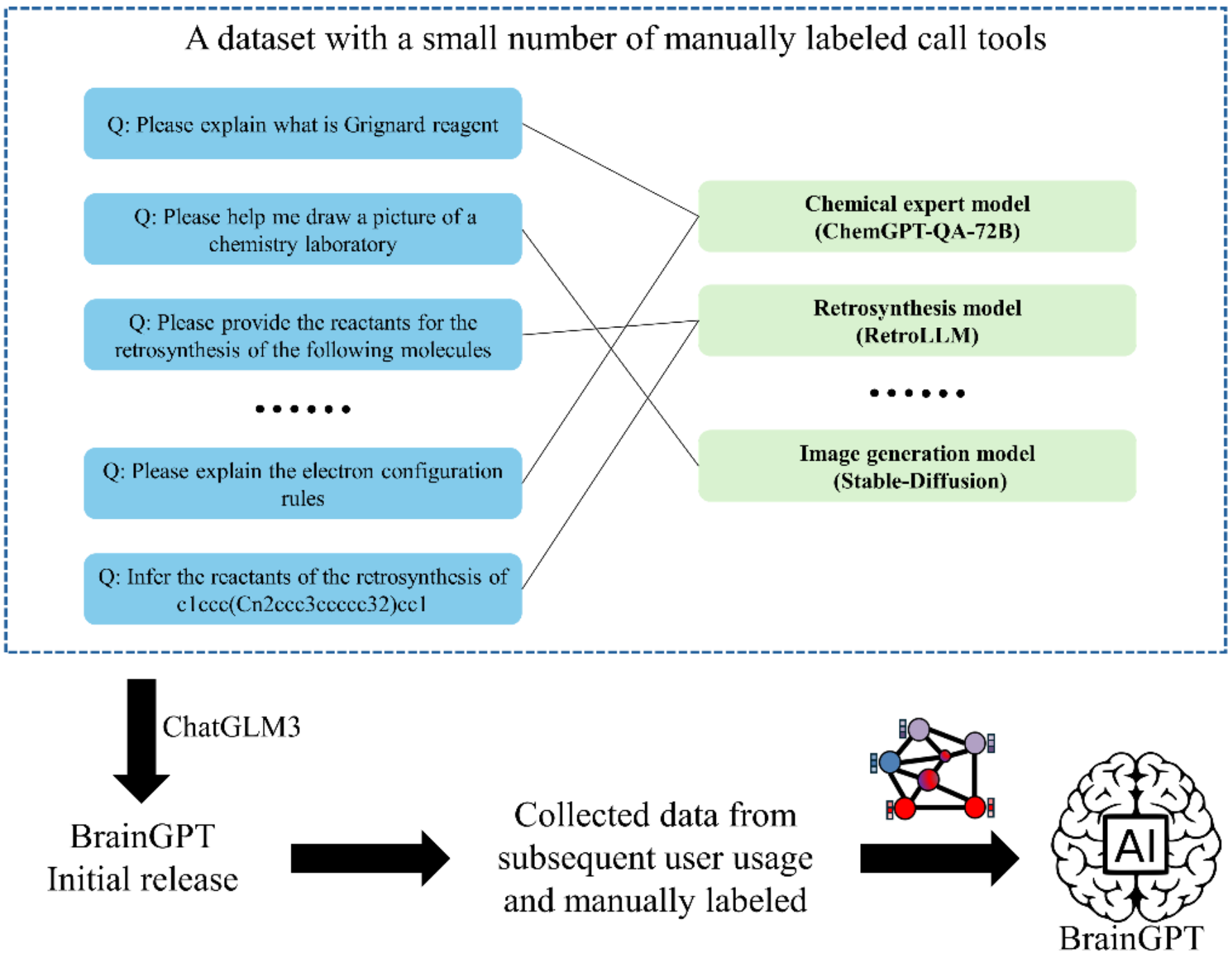

**Figure 7. | Training pipeline of BrainGPT.** The process begins with a dataset with a small number of manually labeled call tools, which are used to fine-tune the ChatGLM3-6B model and produce the initial version of BrainGPT. In the second stage, user-generated dialogue data are continuously collected and annotated to further fine-tune the model. The final version of BrainGPT can accurately interpret chemistry-related questions and intelligently invoke multiple specialized models as needed.

## Additional data and references

Additional data from this study and additional references are provided in the Supplementary Information.

## Corresponding Author

*E-mail: lrhu@chem.ecnu.edu.cn, amzhou@cs.ecnu.edu.cn, xiaohe@phy.ecnu.edu.cn

## Acknowledgments

This paper is dedicated to Professor Youyou Tu, the 2015 Nobel Prize Laureate of Physiology or Medicine on the occasion of her 95th birthday. X.H. was supported by the National Natural Science Foundation of China (Grant Nos. 92477103 and 22273023), Shanghai Municipal Natural Science Foundation (Grant No. 23ZR1418200), the Natural Science Foundation of Chongqing, China (Grant No. CSTB2023NSCQ-MSX0616), the Shanghai Frontiers Science Center of Molecule Intelligent Syntheses, Shanghai Future Discipline Program (Quantum Science and Technology), Shanghai Municipal Education Commission's "Artificial Intelligence-Driven Research Paradigm Reform and Discipline Advancement Program", Guizhou Provincial Science and Technology Projects, China (CXTD 2022001), and the Fundamental Research Funds for the Central Universities. We also acknowledge the Supercomputer Center of East China Normal University (ECNU Multifunctional Platform for Innovation 001) for providing computer resources, and the Syngenta Ph.D. fellowship (We acknowledge Syngenta for the fellowship awarded to Yueqing Zhang).

## Author contributions

A.M.Z., and X.H. conceived the research. Y.Q.Z., W.L. and Y.Z. carried out all the data collection and model training. Y.Q.Z., D.Y.X., J.H.Z., H.H., L.R.H. and X.H. prepared the manuscript. The manuscript was written and revised by all the authors together.

## Data availability

The chemical question-answering test data set in this article comes from OpenCompass[43]. The retrosynthesis training data comes from the 2024_Q1 version of pistachio. The USPTO_50K data set used in the test is an open-source data set. The reaction data collected from the literature is provided in https://github.com/zhangyq9697/ECNU-ChemGPT

## Code availability

The web service for ECNU-ChemGPT is available at https://chemgpt.lrcwtech.com. The BrainGPT model is open-source and available at: https://github.com/1920993165/BrainGPT1. Multi-step retrosynthetic planning was developed and made accessible at https://multiasr.lrcwtech.com/.

## Competing interests

The authors declare no competing interests.

| | LLaMA-3-70B-Instruct | Qwen1.5-72B-Chat | Qwen2-72B-Instruct |
|---|---|---|---|
| **MMLU** | 82.0 | 75.6 | 82.3 |
| **MMLU-Pro** | 56.2 | 51.7 | 64.4 |
| **GSM8K** | 93.0 | 82.7 | 91.1 |
| **MATH** | 50.4 | 42.5 | 59.7 |
| **C-Eval** | 61.6 | 76.1 | 83.8 |

**Table S1.** Comparison of three large language models—LLaMA-3-70B-Instruct, Qwen1.5-72B-Chat, and Qwen2-72B-Instruct—across multiple benchmark tasks, including MMLU, MMLU-Pro, GSM8K, MATH, and C-Eval[1].

| | Data quantity | Data language | Dataset Introduction |
|---|---|---|---|
| C-Eval College chemistry | 24 | Chinese | C-Eval is a comprehensive Chinese evaluation suite for foundation models[2]. |
| C-Eval High school chemistry | 19 | Chinese | |
| C-Eval Middle school chemistry | 20 | Chinese | |
| MMLU College chemistry | 100 | English | MMLU (Massive Multitask Language Understanding) is a new benchmark designed to measure knowledge acquired during pretraining by evaluating models exclusively in zero-shot and few-shot settings. This makes the benchmark more challenging and more similar to how we evaluate humans[3]. |
| MMLU High school chemistry | 203 | English | |
| C_MHChem | 487 | Chinese | C-MHChem-Benchmark-Chinese-Middle-high-school-Chemistry-Test is a High-quality single-choice full-human-writen Benchmark of 600 entries collected from Chinese Chemistry test of middle and high schools past 25 years[4]. |
| GSM8K | 1318 | English | GSM8K is a dataset of 8.5K high quality linguistically diverse grade school math word problems created by human problem writers. The dataset is segmented into 7.5K training problems and 1K test problems[5]. |
| GAOKAO-Bench | 135 | Chinese | This dataset comes from the college entrance examination chemistry test questions collected and sorted in opencompass, mainly subjective questions and open questions[6]. |

**Table S2 All used datasets are introduced in the testing.**

**Supplementary notes 1.** To evaluate the model's performance after fine-tuning, subjective questions were selected for testing, an example of which is shown in Figures S1 and S2. Figure S1 presents a sample question in Chinese along with the standard answer, followed by the responses from ECNU-ChemGPT and Qwen2-72B-instruct. Since ECNU-ChemGPT's output is originally in Markdown format, it was converted to plain text for easier comparison. Figure S2 shows the English version of the question and the standard answer from Figure S1. The English question was then provided to both ECNU-ChemGPT and Qwen2-72B-instruct, and their respective responses were displayed. A comprehensive comparison of the two model's answers reveals that the fine-tuned ECNU-ChemGPT demonstrates a noticeable improvement in the accuracy of chemical knowledge. It can apply chemical principles correctly to analyze problems and provide relevant and accurate responses.

**Question**

10. (15 分) 某班同学用如下实验探究 $\mathrm{Fe}^{2+} 、 \mathrm{Fe}^{3+}$ 的性质. 回答下列问题:
(1) 分别取一定量氯化铁、氯化亚铁固体, 均配制成 $0.1 \mathrm{~mol} / \mathrm{L}$ 的溶液. 在 $\mathrm{FeCl}_{2}$ 溶液中需加入少量铁屑, 其目的是
(2) 甲组同学取 $2 \mathrm{mLFeCl}_{2}$ 溶液. 加入几滴氯水, 再加入 1 滴 $\mathrm{KSCN}$ 溶液, 溶液 变红, 说明 $\mathrm{Cl}_{2}$ 可将 $\mathrm{Fe}^{2+}$ 氧化. $\mathrm{FeCl}_{2}$ 溶液与氯水反应的离子方程式为
(3) 乙组同学认为甲组的实验不够严谨, 该组同学在 $2 \mathrm{mLFeCl}_{2}$ 溶液中先加入 $0.5 \mathrm{~mL}$ 煤油, 再于液面下依次加入几滴氯水和 I 滴 KSCN 溶液, 溶液变红, 煤 油的作用是
(4) 丙组同学取 $10 \mathrm{~mL} 0.1 \mathrm{~mol} / \mathrm{LKI}$ 溶液, 加入 $6 \mathrm{~mL} .1 \mathrm{~mol} / \mathrm{LFeCl}_{3}$ 溶液混合. 分别 取 $2 \mathrm{~mL}$ 此溶液于 3 支试管中进行如下实验:
a.第一支试管中加入 $1 \mathrm{mLCCl}_{4}$ 充分振荡、静置, $\mathrm{CCl}_{4}$ 层呈紫色;
b.第二只试管中加入 1 滴 $K_{3}\left[F e(C N)_{6}\right]$ 溶液，生成蓝色沉淀;
c.第三支试管中加入 1 滴 KSCN 溶液, 溶液变红.实验(2)检验的离子是___ （填离子符号）； 实验(1)和(3)说明： 在 $\mathrm{I}^{-}$过量的情 况下， 溶液中仍含有 (填离子符号), 由此可以证明该氧化还原反应为
(5) 丁组同学向盛有 $\mathrm{H}_{2} \mathrm{O}_{2}$ 溶液的试管中加入几滴酸化的 $\mathrm{FeCl}_{2}$ 溶液, 溶液变成 棕黄色，发生反应的离子方程式为 ； 一段时间后, 溶液中有气泡出现 ， 并放热, 随后有红褐色沉淀生成. 产生气泡的原因是 ； 生成沉淀的 原因是 (用平衡移动原理解释.

**Answer**

(1) 防止氯化亚铁被氧化;
(2) $\mathrm{Cl}_{2}+2 \mathrm{Fe}^{2+}=2 \mathrm{Fe}^{3+}+2 \mathrm{Cl}^{-}$;
(3) 隔离空气（排除氧气对实验的影响）;
(4) $\mathrm{Fe}^{2+}$ ; $\mathrm{Fe}^{3+}$; 可逆反应.
(5) $2 \mathrm{Fe}^{2+}+\mathrm{H}_{2} \mathrm{O}_{2}+2 \mathrm{H}^{+}=2 \mathrm{Fe}^{3+}+2 \mathrm{H}_{2} \mathrm{O}$; 铁离子做过氧化氢分解催化剂分解生成 氧气; 过氧化氢分解反应放热, 促进 $\mathrm{Fe}^{3+}$ 的水解平衡正向移动.

**Answer from ECNU-ChemGPT(Text)**

(1) 在 $FeCl_2$ 溶液中加入少量铁屑的目的是防止 $Fe^{2+}$ 被空气中的氧气氧化。铁屑能与生成的 $Fe^{3+}$ 反应生成 $Fe^{2+}$，从而保持溶液中的 $Fe^{2+}$ 浓度，反应为 $Fe^{3+} + Fe \rightarrow 2Fe^{2+}$。
(2) $FeCl_2$ 溶液与氯水反应的离子方程式为 $2Fe^{2+} + Cl_2 \rightarrow 2Fe^{3+} + 2Cl^-$。
(3) 煤油在实验中的作用是作为隔离层，防止空气中的氧气与溶液接触，避免 $Fe^{2+}$ 被氧化，确保实验结果的准确性。
(4) 实验(b)检验的离子是 $Fe^{2+}$；实验(a)和(c)说明在 $I^-$ 过量的情况下，溶液中仍含有 $Fe^{3+}$，由此可以证明该氧化还原反应为可逆反应。
(5) 向 $H_2O_2$ 溶液中加入几滴酸化的 $FeCl_2$ 溶液，溶液变成棕黄色，发生反应的离子方程式为 $H_2O_2 + 2Fe^{2+} + 2H^+ \rightarrow 2Fe^{3+} + 2H_2O$。一段时间后，溶液中有气泡出现，并放热，随后有红褐色沉淀生成。产生气泡的原因是 $Fe^{3+}$ 催化了 $H_2O_2$ 的分解，生成氧气；生成沉淀的原因是 $Fe^{3+}$ 浓度增加，与溶液中的 $OH^-$ 离子反应生成 $Fe(OH)_3$ 沉淀，即 $3Fe^{3+} + 3H_2O \rightleftharpoons Fe(OH)_3(s) + 3H^+$，由于 $Fe^{3+}$ 浓度增加，平衡向生成沉淀的方向移动。请注意，上述解释是基于化学原理的理论分析。在实际实验中，操作应当严格按照实验室安全规范进行。

**Answer from Qwen2-72B-instruct**

这道题目涉及到铁离子和亚铁离子的化学性质及实验分析，下面我们逐条解析：
(1) 在FeCl2溶液中加入少量铁屑的目的是防止Fe2+被氧化成Fe3+。Fe2+在碱性条件下或者与氧化性较强的物质接触时容易被氧化成Fe3+，铁屑的存在可以提供还原性Fe，通过反应Fe + 2Fe3+ -> 3Fe2+来防止Fe2+的氧化，从而保持FeCl2溶液的稳定。
(2) FeCl2溶液与氯水反应的离子方程式：Cl2 + 2Fe2+ -> 2Cl- + 2Fe3+。在这个反应中，Cl2作为较强的氧化剂，将Fe2+氧化成Fe3+，自身则被还原成Cl-。
(3) 加入煤油的作用主要是隔绝空气，防止Fe2+被空气中的氧气氧化，同时也阻止其它可能的副反应，保持实验条件的纯净，确保Cl2对Fe2+的氧化是唯一反应。
(4) 实验(b)中加入1滴K3[Fe(CN)6]溶液产生蓝色沉淀，这是特用来检测Fe2+的K3[Fe(CN)6]试剂反应，说明溶液中还存在Fe2+。实验表明即便在I-过量的情况下，溶液中仍含有Fe3+，这说明Fe3+与I-的反应是一个可逆反应。
(5) 向H2O2溶液中加入酸化的FeCl2溶液时，首先发生H2O2 + 2Fe2+ + 2H+ -> 2H2O + 2Fe3+，Fe2+作为催化剂将H2O2分解成水和氧气，同时自身被氧化成Fe3+。一段时间后，溶液中出现气泡是由于H2O2分解产生的氧气。随后溶液中Fe3+的浓度增加，其与水反应形成红褐色的Fe(OH)3沉淀。生成沉淀的原理解释是酸性减弱，导致Fe3+与水反应生成Fe(OH)3沉淀的趋势增加，即Fe3+ + 3H2O ⇌ Fe(OH)3(s) + 3H+，根据勒夏特列原理，H+的减少促使平衡正向移动，导致Fe(OH)3沉淀的生成。综上，该题目的理论分析与实验现象一致，涉及了Fe2+和Fe3+的相互转化、氧化还原反应及沉淀反应等内容。

**Figure S1.** An example from the OpenCompass college entrance examination dataset, including questions, standard answers, ECNU-ChemGPT's answers, and Qwen2's answers.

**Question**

10. (15 points) Students in a class use the following experiment to explore the properties of $\mathrm{Fe}^{2+} and \mathrm{Fe}^{3+}S. Answer the following questions:

(1) Take a certain amount of ferric chloride and ferrous chloride solids, and prepare them into $0.1 \mathrm{~mol} / \mathrm{L}S solutions. A small amount of iron filings needs to be added to the S\mathrm{FeCl}_{2}$ solution.

(2) The purpose is to Group A students take $2 \mathrm{mLFeCl}_{2}$ solution. Add a few drops of chlorine water, and then add 1 drop of $\mathrm{KSCN}S solution. The solution turns red, indicating that $\mathrm{Cl}_{2}S can oxidize S\mathrm{Fe}^{2+}$. S\mathrm{FeCl}_{2}$ The ionic equation of the reaction between the solution and chlorine water is .

(3) Group B students thought that Group A's experiment was not rigorous enough. They first added $0.5 \mathrm{~mL}$ kerosene to $2 \mathrm{mLFeCl}_{2}$ solution, and then added a few drops of chlorine water and 1 drop of KSCN solution below the liquid surface. The solution turned red. The role of kerosene is .

(4) Group C students took $10 \mathrm{~mL} 0.1 \mathrm{~mol} / \mathrm{LKI}$ solution, added $6 \mathrm{~mL} .1 \mathrm{~mol} / \mathrm{LFeCl}_{3}$ solution and mixed it. Take $2 \mathrm{~mL}$ of this solution in 3 test tubes and conduct the following experiments:

a. Add $1 \mathrm{mLCCl}_{4}$ to the first test tube, shake it thoroughly, and let it stand. $\mathrm{CCl}_{4}S layer is purple;

b. Add 1 drop of $K_{3}\left[Fe(C N)_{6}\right]$ solution to the second test tube, and a blue precipitate is generated;

c. Add 1 drop of KSCN solution to the third test tube, and the solution turns red. The ion tested in experiment (2) is ___ (fill in the ion symbol); Experiments (1) and (3) explain: When S\mathrm{I}^{-}$ is in excess, the solution still contains (fill in the ion symbol), which proves that the redox reaction is .

(5) Group D students added a few drops of acidified $\mathrm{FeCl}_{2}$ solution to the test tube containing $\mathrm{H}_{2} \mathrm{O}_{2}S solution, and the solution turned brown-yellow. The ionic equation of the reaction is; after a period of time, bubbles appeared in the solution and released heat, Then a reddish brown precipitate is formed. The reason for the bubbles is ; the reason for the precipitate is (explained by the principle of equilibrium shift.

**Answer**

(1) Prevent ferrous chloride from being oxidized;

(2) S\mathrm{Cl}_{2}+2 \mathrm{Fe}^{2+}=2 \mathrm{Fe}^{3+}+2 \mathrm{Cl}^{-}$;

(3) Isolate air (eliminate the influence of oxygen on the experiment);

(4) S\mathrm{Fe}^{2+} ; \mathrm{Fe}^{3+}$; Reversible reaction.

(5) S2 \mathrm{Fe}^{2+}+\mathrm{H}_{2} \mathrm{O}_{2}+2 \mathrm{H}^{+}=2 \mathrm{Fe}^{3+}+2 \mathrm{H}_{2} \mathrm{O}$; Iron ions decompose to generate oxygen through hydrogen peroxide decomposition catalyst; hydrogen peroxide decomposition reaction releases heat, promoting the positive movement of the hydrolysis equilibrium of $\mathrm{Fe}^{3+}$.

**Answer from ECNU-ChemGPT(Text)**

1.Purpose of Adding Iron Filings : The purpose of adding a small amount of iron filings to the $FeCl_2$ solution is to ensure that any $Fe^{3+}$ present is reduced back to $Fe^{2+}$ , maintaining the solution's composition as $Fe^{2+}$ . This is done using the half-reaction: $Fe+2Fe^{3+}\rightarrow 3Fe^{2+}$ .

2.Ionic Equation for Group A's Experiment : The ionic equation for the reaction of $FeCl_2$ solution with chlorine water is: $2Fe^{2+}+Cl_2\rightarrow 2Fe^{3+}+2Cl^-$ .

3.Role of Kerosene in Group B's Experiment : Kerosene is added to create a physical barrier between the solution and the air. It serves to prevent atmospheric oxygen from oxidizing the $Fe^{2+}$ ions, ensuring that any observed oxidation is solely due to the reaction with chlorine water.

4.Group C Experiments : Experiment (a) : The purple color in the $CCl_4$ layer indicates the presence of iodine ($I_2$ ), suggesting that $Fe^{3+}$ ions have oxidized $I^-$ to $I_2$. Experiment (b) : The blue precipitate formed with the addition of $K_3[Fe(CN)_6]$ is ferric hexacyanoferrate(II), indicating the presence of $Fe^{2+}$ ions. Experiment (c) : The red color is due to the formation of the FeSCN2+ complex, indicating the presence of $Fe^{3+}$ . Ions tested in experiment (2): $Fe^{2+}$ ions. Experiments (1) and (3) explain that when I– is in excess, the solution still contains $Fe^{3+}$ ions (as shown by the red color in experiment (c)), which proves that the redox reaction between $Fe^{3+}$ and $I^-$ is not complete, indicating a reversible reaction.

5.Group D's Experiment : Ionic Equation : The reaction between $Fe^{2+}$ and $H_2O_2$ in an acidified solution is: $2Fe^{2+}+2H^++H_2O_2\rightarrow 2Fe^{3+}+2H_2O$ . Reason for the Bubbles : The appearance of bubbles is due to the decomposition of $H_2O_2$ , catalyzed by $Fe^{3+}$ , generating oxygen gas ($O_2$ ). Reason for the Formation of the Precipitate : The formation of the reddish-brown precipitate is attributed to the hydrolysis of $Fe^{3+}$ ions, forming $Fe(OH)_3$ . This occurs due to the shift in equilibrium, as the increased concentration of $Fe^{3+}$ ions favors the formation of the less soluble $Fe(OH)_3$ precipitate according to Le Chatelier's principle. These experiments effectively demonstrate various redox reactions involving $Fe^{2+}$ and $Fe^{3+}$ , as well as the principles of equilibrium and complex formation in aqueous solutions.

**Answer from Qwen2-72B-instruct**

The students' experiments involve exploring the redox properties of iron compounds, particularly $Fe^{2+}$ and $Fe^{3+}$ . Let's answer the questions and explain the various parts of the experiments step by step.

1.Purpose of Adding Iron Filings : The purpose of adding iron filings to the $FeCl_2$ solution is not explicitly stated, but it might be inferred that the iron filings could be used to reduce any $Fe^{3+}$ present to $Fe^{2+}$ , keeping the solution as a pure $Fe^{2+}$ solution.

2.Group A's Experiment : Group A students added a few drops of chlorine water (which contains $Cl_2$ ) to a $FeCl_2$ solution. After adding KSCN solution, the solution turns red, indicating the presence of $Fe^{3+}$ . The ionic equation for the reaction between $FeCl_2$ and chlorine water is: $2Fe^{2+}+Cl_2\rightarrow 2Fe^{3+}+2Cl^-$

3.Group B's Experiment : Group B students questioned the rigor of Group A's experiment and introduced a layer of kerosene on top of the $FeCl_2$ solution before adding chlorine water and KSCN . The role of kerosene is to prevent direct exposure to air, thereby ensuring that any oxidation of $Fe^{2+}$ to $Fe^{3+}$ is solely due to the chlorine water added, not atmospheric oxygen.

4.Group C's Experiment : Group C students mixed a KI solution with a $FeCl_3$ solution and analyzed the products using three different methods. a. $CCl_4$ Layer Turning Purple : The $CCl_4$ layer turning purple indicates extraction of iodine from the aqueous layer, confirming the oxidation of $I^-$ to $I_2$ . b. Blue Precipitate Formation : The blue precipitate formed on adding a $K_3[Fe(CN)_6]$ drop indicates the presence of $Fe^{2+}$ . The ion tested is $Fe^{2+}$. c. Solution Turning Red : The solution turning red upon adding KSCN solution after $FeCl_3$ had been added suggests the presence of $Fe^{3+}$ . This demonstrates that even when I– is in excess, $Fe^{3+}$ remains in the solution, proving the redox reaction is not quantitative, i.e., it does not go to completion.

5.Group D's Experiment : Group D students observed changes in a solution after adding a few drops of acidified $FeCl_2$ solution to a $H_2O_2$ solution. The ionic equation for this reaction (also called the Fenton's Reaction) is: $2Fe^{2+}+H_2O_2+2H^+\rightarrow 2Fe^{3+}+2H_2O$. After a while, bubbles appeared, indicating the release of oxygen gas as a result of the decomposition of $H_2O_2$ catalyzed by $Fe^{3+}$. The reddish-brown precipitate is likely $Fe(OH)_3$ , which forms due to the shift in equilibrium when the concentration of $Fe^{3+}$ increases. With an excess of water and possibly some additional hydroxide ions from water's autoprotolysis, the equilibrium shifts towards the formation of $Fe(OH)_3$ , which is a solid. The Le Chatelier's principle explains this shift—since the concentration of $Fe^{3+}$ is increased by the reaction, the equilibrium will shift to reduce this increase by producing more of the solid $Fe(OH)_3$ .

**Figure S2.** An example from the OpenCompass college entrance examination dataset, including questions, standard answers, and after translating them into English, the answers from ECNU-ChemGPT and Qwen2 are obtained.

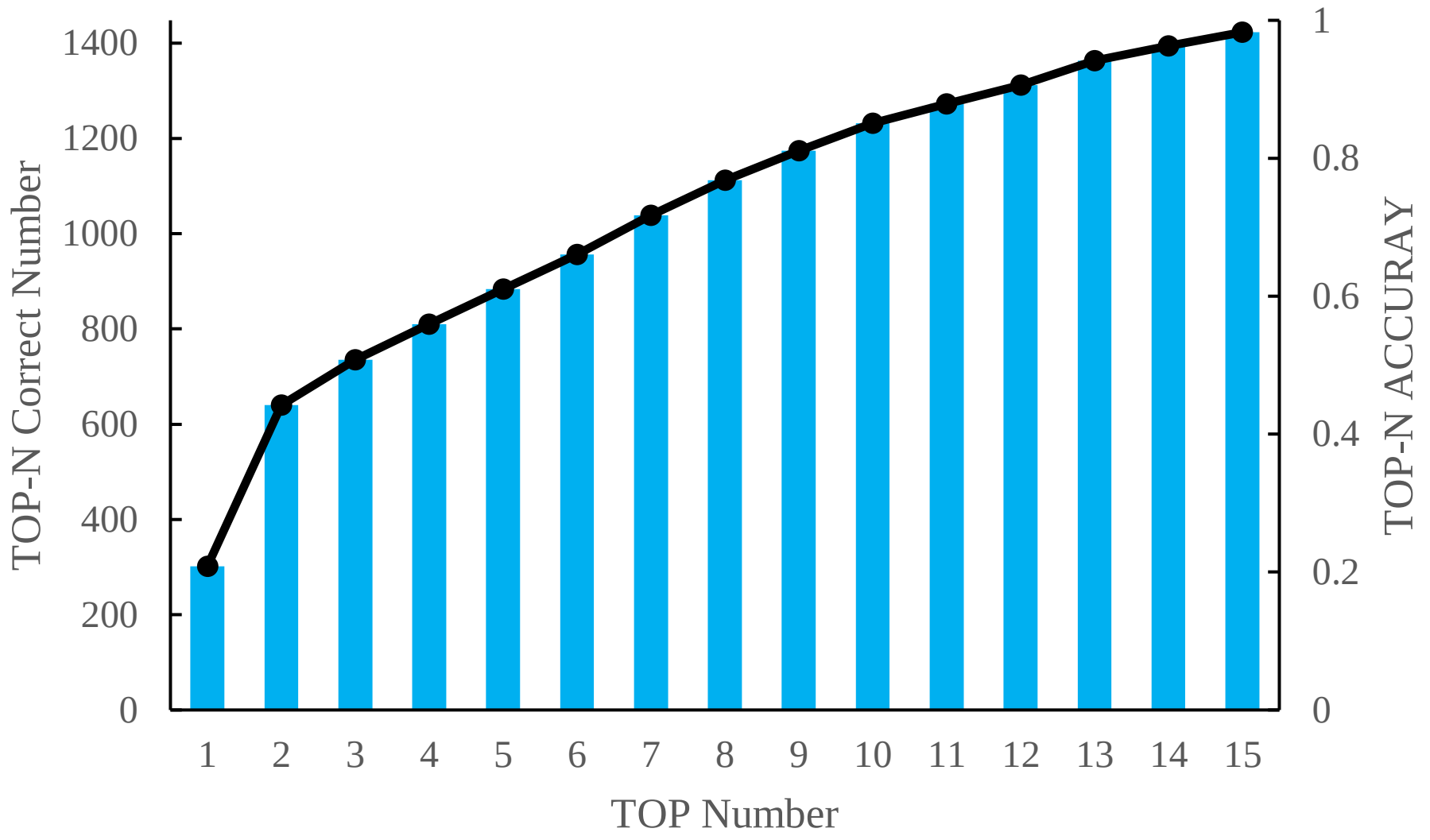

**Figure S3.** A bar chart of the proportion of model inference results with a similarity of 95% or more to the experimental results. Under this relaxed standard, the accuracy of Top-1, Top-3, Top-10, and Top-15 increased to 20.9%, 50.8%, 85.1%, and 98.2%, respectively.

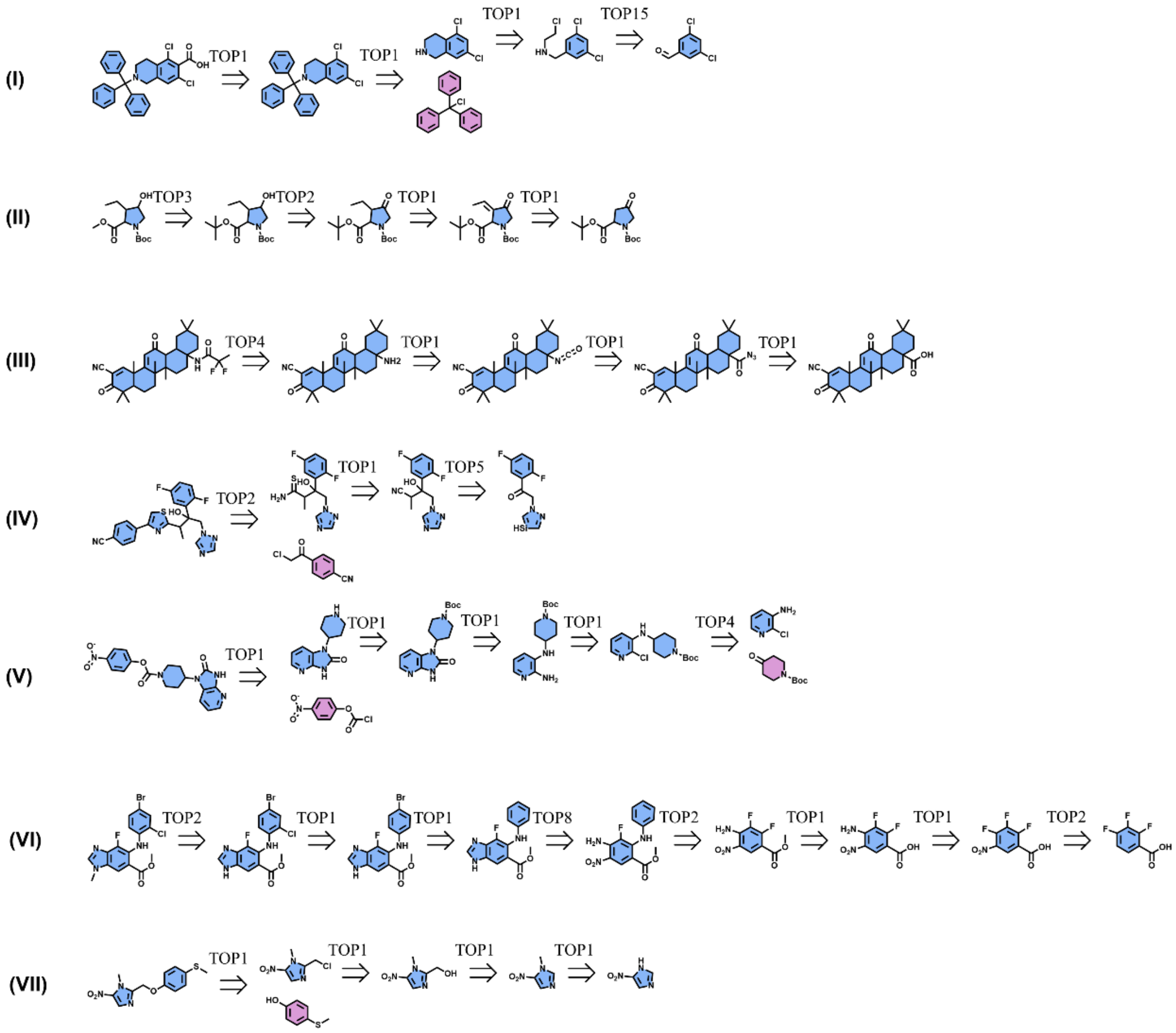

**Figure S4.** Predicted synthesis routes exactly matched the experimental pathways. In the figure, routes 1 to 7 are the synthetic routes of Tetrahydroisoquinoline-6-carboxylic Acid Fragment for Lifitegrast[7], Voxilaprevir Pyrrolidino[8], Omaveloxolone[9], Isavuconazole[10], Rimegepant Intermediate Carbamate[11], Selumitinib Benzimidazole[12], Fexinidazole[13].

**Table S2 Parameter Settings for the Training Stage**

```
torchrun $DISTRIBUTED_ARGS finetune.py \
    --model_name_or_path $MODEL \
    --data_path $DATA \
    --bf16 True \
    --output_dir $OUTPUT_PATH \
    --num_train_epochs 1 \
    --per_device_train_batch_size 4 \
    --per_device_eval_batch_size 1 \
    --gradient_accumulation_steps 1 \
    --evaluation_strategy "no" \
    --save_strategy "steps" \
    --save_steps 500 \
    --save_total_limit 10 \
    --learning_rate 3e-5 \
    --weight_decay 0.1 \
    --adam_beta2 0.95 \
    --warmup_ratio 0.01 \
    --lr_scheduler_type "cosine" \
    --logging_steps 1 \
    --report_to "none" \
    --model_max_length 4096 \
    --lazy_preprocess True \
    --use_lora True\
    --gradient_checkpointing \
    --deepspeed $[1]
```

**Table S3 Parameter Settings for the LoRA**

```
lora_r: 64
lora_alpha: 16
lora_dropout: 0.05
lora_target_modules: ["q_proj", "k_proj", "v_proj", "o_proj", "gate_proj", 'up_proj', 'down_proj', 'embed_tokens', 'lm_head'])
q_lora: False
```

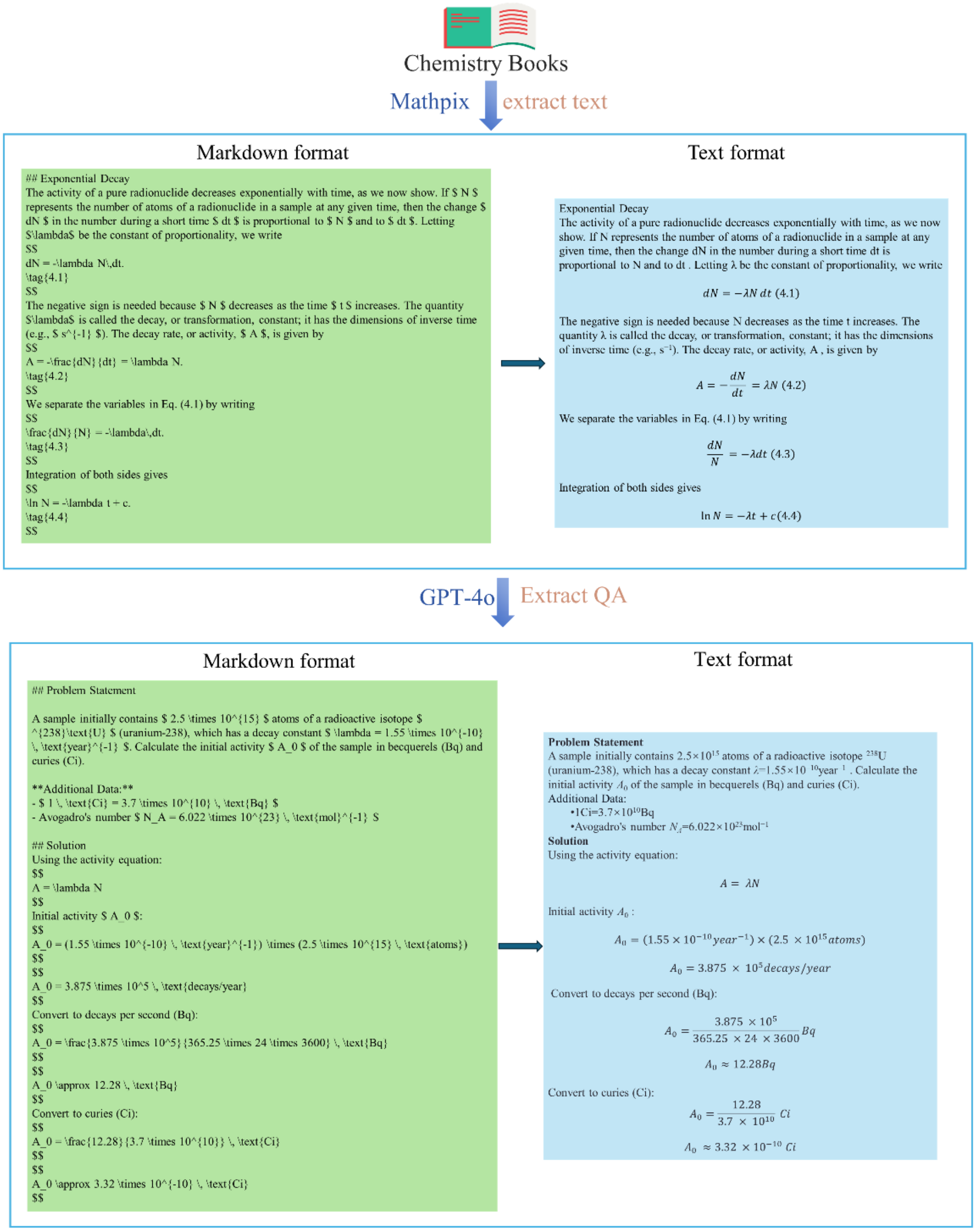

**Figure S5.** The figure shows an example of extracting markdown format text from a book and using LLM to obtain a high-quality Q&A dataset. First, Mathpix is used to extract the knowledge content in the Markdown format from the book, and then the LLM API is used to convert the text into a Q&A dataset. The green part in the figure is the content in the Markdown format, and the blue part is the content converted to Text format for human viewing.

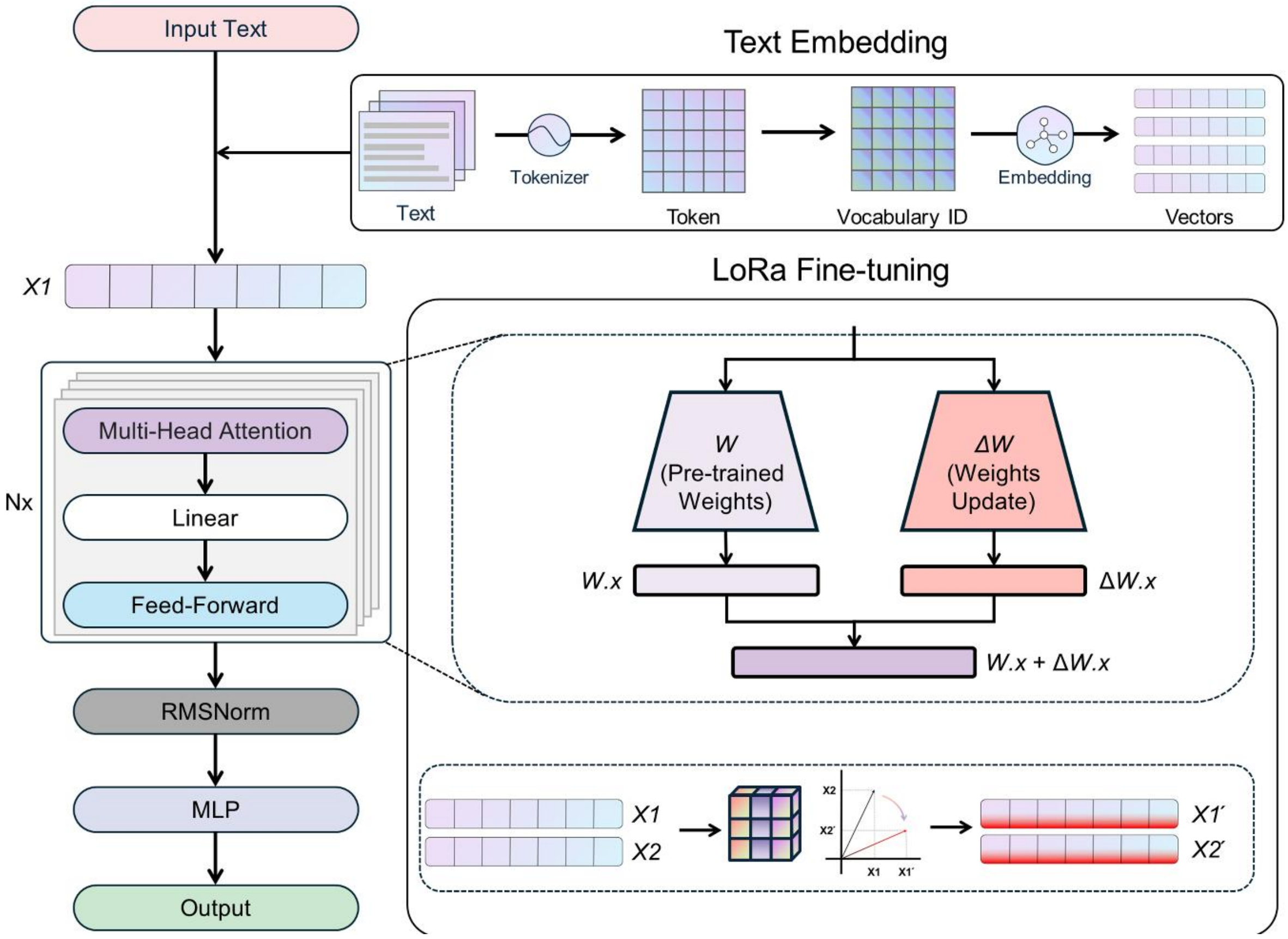

**Figure S6.** Base model architecture and LoRA fine-tuning schematics. In order to reduce the parameter scale and computational cost of large language models when fine-tuning downstream tasks, this paper adopts the LoRA (Low-Rank Adaptation) method for efficient parameter fine-tuning. The core idea of LoRA is to freeze part of the weight matrix in the pre-trained model (usually the Query and Value weights in the attention mechanism), and introduce two low-rank trainable matrices to reconstruct parameter updates, thereby achieving performance improvement without modifying the original model weights. Compared with full parameter fine-tuning, LoRA significantly reduces the video memory and the number of trainable parameters required for training, and is suitable for model customization in resource-constrained environments.